\documentclass[a4paper,11pt]{article}

\usepackage[a4paper,
  left=2.5cm, right=2.5cm,
  top= 3cm, bottom=4cm]{geometry}
\usepackage{amsmath}
\usepackage{oldgerm}
\usepackage{amssymb}
\usepackage{bbm}
\usepackage[dvips]{graphicx}
\usepackage{epsfig}
\usepackage{color}
\usepackage{cite}
\usepackage{epic}
\usepackage{hyperref}

\newcommand{\be}{\begin{equation}}  
\newcommand{\ee}{\end{equation}}

\newcommand{\rem}[1]{} 

\def\C{\mathbb{C}}
\def\Z{\mathbb{Z}}
\def\R{\mathbb{R}}

\def\P{\mathbb{P}}
\def\cO{\mathcal{O}}
\def\I{\mathbb{I}}

\def\Hirz[#1]{\mathbbm{F}_{#1}}
\def\o[#1]{\overline{#1}}

\def\cL{\mathcal{L}}

\def\cD{\mathcal{D}}

\title{The fate of ${\rm U}(1)$'s at strong coupling in F-theory}

\begin{document}
\begin{titlepage}
 
\vskip -0.5cm
\rightline{\small{\tt KCL-MTH-14-02}} 
 
\begin{flushright}

\end{flushright}
 
\vskip 1cm
\begin{center}
 
{\Large \bf The fate of ${\rm U}(1)$'s at strong coupling in F-theory} 
 
 \vskip 1.2cm
 
 A.~P.~Braun$^1$, A.~Collinucci$^2$ and R.~Valandro$^{3}$

 \vskip 0.4cm
 
 {\it $^1$Department of Mathematics, King's College, London WC2R 2LS, UK
\\[2mm]

 $^2$Physique Th\'eorique et Math\'ematique, Universit\'e Libre de Bruxelles, C.P. 231, \\1050
Bruxelles, Belgium \\[2mm]
 
 $^3$ICTP, Strada Costiera 11, Trieste 34014, Italy.
 }
 \vskip 1.5cm
 
\abstract{${\rm U}(1)$ gauge symmetries in F-theory are expected to manifest themselves as codimension three singularities of Calabi-Yau fourfolds. However, some of these are known to become massive at strong coupling via the St\"uckelberg mechanism.
In this note, we propose a geometric picture for detecting all ${\rm U}(1)$'s, and determining which ones are massive and which ones are massless. We find that massive gauge symmetries show up as codimension three singularities that only admit small, non-K\"ahler, resolutions. Our proposal passes several highly non-trivial tests, including a case with a non-diagonal mass matrix.} 

\end{center}


\end{titlepage}

\tableofcontents

\section{Introduction}
F-theory is a powerful tool for studying holomorphic quantities of IIB string theory in the strong coupling regime. By geometrizing the data of the 7-branes, and via its duality with M-theory, F-theory brings the Dynkin diagrams representing gauge symmetries to life by literally producing them inside elliptically fibered Calabi-Yau manifolds.
Indeed, the correspondence between gauge groups and geometry has been understood for a long time \cite{Bershadsky:1996nh}, \emph{up to Abelian factors}.

Surprisingly, Abelian factors of the gauge group are the most difficult to `see' in the geometry of the compactification. Moreover, some of these might acquire mass by St\"uckelberg mechanism, 
turning them into global symmetries. In the weakly coupled IIB picture, which contains only O7-planes and D7-branes, the following rule was established \cite{Ibanez:1998qp,Poppitz:1998dj,Aldazabal:2000dg,Plauschinn:2008yd,Grimm:2010ez,Grimm:2011dj}

{\it Consider a stack of $N$ D7-branes with unitary gauge group (i.e. not invariant under the orientifold involution $\sigma$) wrapping a complex surface of class $[D7]$, and an orientifold image stack wrapping a surface of class $\sigma^*[D7]$. Then, the diagonal ${\rm U}(1) \subset U(N)$ remains massless at $g_s \neq 0$ if and only if $[D7] = \sigma^*[D7]$ as homology classes of the compactification threefold.
}

This statement begs the following question: {\it Is it possible to detect ${\rm U}(1)$'s directly in an F-theory CY fourfold and to discriminate between the massive and massless ones?} In this paper, we provide an answer to this question. 

Current understanding of F-theory dictates that ${\rm U}(1)$ gauge symmetries manifest themselves as codimension three singularities in the fourfold.
Intuitively, suppose that the Weierstrass model for the F-theory CY fourfold has the form $A\,B = C\,D$, for four generic polynomials. Then, the fourfold has a codimension three singularity, a one-parameter family of conifolds, which admits a small, K\"ahler resolution. The resolved fourfold will have two new Cartier divisors $D_{\pm}$, whose ancestors are the ideals\footnote{Throughout this paper, instead of defining a submanifold as $f_1= 0 \cap f_2=0 \cap \ldots \cap f_n=0$, we will simply denote its ideal as $(f_1, f_2, \ldots, f_n)$. } $(A,\,C)$ and $(A\,, D)$. Such new divisors will imply the existence of two new closed two-forms $\omega_\pm \in H^{1,1}($CY$_4$). The 11d SUGRA three-form $C_3$ can then be reduced along the 4d Poincar\'e invariant combination $\omega_+-\omega_-$ with the Ansatz $C_3 = (\omega_+-\omega_-) \wedge A_\mu$, where $A_\mu$ is a photon in the 3d and the T-dual 4d effective theories.
Hence, the link between codimension three singularities and ${\rm U}(1)$'s.

This cannot be the whole story about ${\rm U}(1)$'s, since we know that there must also be massive photons, even in absence of background fluxes. 
In type IIB, if a D7-brane and its image lie in different homology classes, a mass is generated for the corresponding ${\rm U}(1)$ by the geometric St\"uckelberg mechanism \cite{Ibanez:1998qp,Poppitz:1998dj,Aldazabal:2000dg,Plauschinn:2008yd,Grimm:2010ez,Grimm:2011dj,Grimm:2011tb}.
The responsible coupling is given by the following Chern-Simons term in the 4d effective action \cite{Jockers:2004yj}
\begin{equation}\label{GeomStuckMass}
 S_{\rm Stck} \sim \sum_{a} n^a \int_{\mathbb{R}^{1,3}} F_2 \wedge c_2^a  
\end{equation}
where $n^a$ are the coefficients in the expansion $[D7]=n^\alpha \cD_\alpha+n^a \cD_a$, with $\cD_\alpha$ a basis of even 2-forms and $\cD_a$ a basis of odd 2-forms, $F_2$ is the 4d field strength on $D7_\ell$ and $c_2^a$ are the 4d two-form dual to the axions coming from the reduction of the RR two-form potential $C_2$. The generated mass is proportional to the string coupling $g_s$. 

In \cite{Grimm:2010ez} (and subsequently in \cite{Grimm:2011tb,Grimm:2013fua}) it was predicted that a massive photon $A_{IIB}$ on a D7-stack in IIB should lift to a $C_3$-form in M-theory that can be decomposed as $A_{IIB} \wedge \omega_{nh}$, where $\omega_{nh}$ is a non-harmonic two-form. Some clues were provided as to what the mechanism at play might be, but a clear geometric picture has remained elusive until now.

In order to answer the question, it is crucial to understand how to take the weak coupling limit of F-theory. Sen's limit \cite{Sen:1997gv, Sen:1997kw, Sen:1996vd} consists in taking a particular one-parameter family of CY fourfolds, expanding the discriminant of the elliptic fibration in terms of the parameter $\epsilon$, and keeping the leading term. Although this approach has proven to be very fruitful, it is limited by the fact that the limit $\epsilon \rightarrow 0$ can only be taken \emph{after} computing the discriminant. Doing it directly in the fourfold would brutally mutilate the geometry, washing away most of the 7-brane data.
In other words, we do not have a `weakly coupled F-theory fourfold' \emph{per se}.

Recently, the Sen limit was re-conceived in \cite{Clingher:2012rg} in a way that completely addresses the issue. The basic idea is to consider the whole one-parameter family of CY fourfolds, which is itself a CY five-fold, and blowing up the singular locus inside the \emph{central fiber} at $\epsilon=0$. In this way, a new fourfold emerges that has two components. One component only sees perturbative physics, i.e. the only monodromies in the fiber are remnants of $T$ and $-\I_2 \subset SL(2, \Z)$. 

This clean way of geometrizing Sen's limit allows us to track the fate of a ${\rm U}(1)$ gauge group as it undergoes its transition from the perturbative to the non-perturbative regime. Our results can be summarized as follows:

{\it A ${\rm U}(1)$ gauge group gives rise to a codimension three singularity in the CY fourfold, provided there exists matter charged under this group. The singularity can always be seen as a family of conifold singularities fibered over the charged matter curves in IIB. A ${\rm U}(1)$ remains massless at strong coupling if and only if the corresponding singularity admits a small, K\"ahler resolution.}

This paper is organized as follows: In section \ref{sec:CDW}, we introduce Sen's limit as redefined by Clingher, Donagi and Wijnholt, which will serve as our framework. Sections \ref{sec:brane/imagebrane} and \ref{sec:examples2} display the simplest examples of: a ${\rm U}(1)$ that remains massless, and one that acquires a St\"uckelberg mass at $g_s \neq 0$, respectively. In the latter, we see a curve worth of conifold singularities admitting a small K\"ahler resolution at weak coupling, which becomes non-K\"ahler at strong coupling. 

Section \ref{sec:globalexamples} contains three global models of increasing levels of complexity, putting our geometric picture to stringent tests. The first two consist of a brane/image-brane pair with a tadpole saturating, invariant Whitney brane. In the first one, we detect a massive ${\rm U}(1)$, whereas in the second one, a massless ${\rm U}(1)$. 

In the third and most intricate model, we have two brane/image-brane pairs. The first pair has a massless ${\rm U}(1)$, but no charged matter under it. The second pair has a massive ${\rm U}(1)$. Naively, we shouldn't see any massless ${\rm U}(1)$ manifestly in the strongly coupled fourfold. However, the two by two mass matrix still predicts that a massless linear combination of ${\rm U}(1)$'s survives, with matter charged under it. 
We find that the singularity structure of the CY fourfold perfectly reflects this behavior, thereby providing strong support for our proposal.

\section{The weak coupling limit geometrized} \label{sec:CDW}
Let us first establish some definitions to set up the notation. We start from a smooth Calabi-Yau fourfold $X_4$ that is an elliptic fibration over the base manifold $B_3$, with a section. We  describe this manifold by the Weierstrass model
\begin{equation} \label{WM}
y^2=x^3 + x z^4 f+z^6 g \:,
\end{equation}
i.e. as a hypersurface equation in the ambient five-fold $X_5=\P_{2,3,1}(\cO_{B_3}\oplus\cO_{B_3} \oplus K )$. Hence, $x,y,z$ are sections respectively of $F^{\otimes 2}$,  $F^{\otimes 3}$ and  $F\otimes K$ where $F$ is the line bundle associated with the projective action in the fiber direction and $K$ is the canonical bundle of the base manifold 
$B_3$. In equation \eqref{WM}, $f,g$ are sections of $\bar{K}^{\otimes 4},\bar{K}^{\otimes 6}$. They can be expressed in terms of the sections 
$a_i\in \bar{K}^{\otimes i}$ appearing in the Tate form of the Weierstrass model:
\begin{equation}
 \begin{array}{l}
 f=-\frac{1}{48}\left(b_2^2-24b_4\right) \\ \\
g=\frac{1}{864}\left(b_2^3-36b_2b_4+216b_6\right) \\
 \end{array} \qquad \qquad \mbox{where} \qquad \qquad
 \begin{array}{l} \label{definitionbi}
 b_2=a_1^2+4a_2 \\
b_4=a_1a_3+2a_4 \\
b_6=a_3^2+4a_6 
 \end{array} \:.
\end{equation}

Sen's weak coupling as described in \cite{Clingher:2012rg,Donagi:2012ts} consists in taking $f$ and $g$ given by \eqref{definitionbi} and scaling the sections $b_4$ and $b_6$ as 
\begin{equation} \label{wcscaling}
 b_4 \rightarrow \epsilon \, b_4 \qquad \qquad b_6 \rightarrow \epsilon^2 \, b_6 \:,
\end{equation}
where $\epsilon \subset \C$. As $\epsilon \rightarrow 0$, the string coupling becomes small almost everywhere on $B_3$. 
By introducing the coordinate $s=x-\frac{1}{12}b_2z^2$, and using the parametrization \eqref{wcscaling} for $f$ and $g$, one obtains (after rescaling the $b_i$'s by a factor of four)
\begin{equation}
 y^2 = s^3 + b_2 s^2\, z^2+ 2b_4 s\,\epsilon\,z^4 + b_6 \epsilon^2\,z^6 \:.
\end{equation}
This equation describes a family of Calabi-Yau fourfolds over the $\epsilon$-plane. At $\epsilon=0$ the fourfold has a severe degeneration. In particular the elliptic fiber degenerates over all the points in the base manifold $B_3$ and the information on the D7-brane locus is lost.

As explained in \cite{Clingher:2012rg,Donagi:2012ts}, the key to taming this degeneration is to consider the whole one-parameter family of fourfolds as a CY fivefold in its own right. One then performs a blow-up of said fivefold at the singular locus $y=s=\epsilon=0$. This is accomplished by blowing up the ambient space via the introduction of a new coordinate $\lambda$, with the following projective $\C^*$ actions:

\begin{center}
\begin{tabular}{ccccc}
$y$ & $s$ & $t$ & $z$ & $\lambda$ \\
$3$ & $2$ & $0$ & $1$ & $0$ \\
$1$ & $1$ & $1$ & $0$ & $-1$ \\
$3\bar{K}_B$ & $2\bar{K}_B$ & $0$ & $0$&$0$
\end{tabular}\\
\end{center}

The SR ideal is generated by $[syz],[yst],[z\lambda]$. The blow-down map is
$s\mapsto s \lambda$, $y\mapsto y\lambda$, $\epsilon \mapsto t\lambda$.

The blown-up fivefold is described as the following hypersurface inside this ambient space:
\begin{equation} \label{completecentralfiber}
W_5:\quad y^2=s^3\lambda + z^2\left(\begin{array}{cc}s & tz^2 \end{array}\right) 
\left(\begin{array}{cc}b_2 & b_4 \\ b_4 & b_6 \end{array}\right) \left(\begin{array}{c}s \\ tz^2 \end{array}\right)
\end{equation}

The fourfolds over $\epsilon \neq 0$ are isomorphic to the original CY fourfold. However, the fourfold over the \emph{central fiber} at $\epsilon=0$, which represents the weak coupling limit, splits up into two components: $\epsilon = t\lambda=0$. 

In summary, the central fiber at $\epsilon=0$ looks like $W_T \cup_{X_3} W_E$. Here
\begin{align}
 W_T:\quad W_5\cap\{t=0\}:\quad & y^2=s^2(b_2z^2+s\lambda)\\
 W_E:\quad W_5\cap\{\lambda=0\}:\quad &   y^2=z^2\left(\begin{array}{cc}s & tz^2 \end{array}\right) 
 \left(\begin{array}{cc}b_2 & b_4 \\ b_4 & b_6 \end{array}\right) \left(\begin{array}{c}s \\ tz^2 \end{array}\right)
\end{align}

The two components $W_E$ and $W_T$ intersect in a space $X_3$ which can be indentified with the double cover Calabi-Yau threefold where the perturbative IIB string theory lives:
\begin{equation}
 X_3:\quad W_5\cap \{t=0\}\cap\{\lambda=0\}:\quad y^2=z^2s^2 b_2 \, .
\end{equation}
Note that, due to the SR ideal, the divisors $[z]$ and $[s]$ do not meet $X_3$, so we can define
\begin{equation}
 \xi=y/(sz) \, 
\end{equation}
and write $X_3$ in the standard way as 
\begin{equation}\label{X_3}
 \xi^2=b_2 \, .
\end{equation}

In the following, we will be interested in $W_E$, defined by $\lambda=0$. The SR-ideal indicates that we can set $z=1$. $W_E$ is then given by the hypersurface equation
\begin{align}\label{WEeq}
 W_E:\quad &   y^2= b_2s^2+2b_4s\,t+b_6 t^2 
\end{align}
in an ambient five-fold $Y_5$, that is spanned by the base manifold $B_3$ and the homogeneous coordinates $(s:y:t)$ with weights:

\begin{equation}
\begin{tabular}{ccc}
$y$ & $s$ & $t$ \\
$1$ & $1$ & $1$ \\
$3\bar{K}_B$ & $2\bar{K}_B$ & $0$
\end{tabular}
\end{equation}

Hence, \eqref{WEeq} is a bundle of quadratic equations in $\P^2$ (a conic bundle) that describes a $\P^1$ fibration over $B_3$. The $\P^1$ fiber degenerates into two $\P^1$'s over the discriminant locus of the quadric
\begin{equation}
\Delta_E\equiv  \det\left(\begin{array}{cc}b_2 & b_4 \\ b_4 & b_6 \end{array}\right)=0 \quad\mbox{in} \, B_3\,.
\end{equation}
This is the locus of the $D7$-brane in $B_3$.
The type IIB Calabi-Yau threefold sits inside $W_E$ at the locus $t=0$. This is given by two points on the $\P^1$ fiber, that are exchanged when going around the locus $b_2=0$ on the base $B_3$.

Note that the $\P^1$-fiber can be considered as a piece of the elliptic fiber of the full CY fourfold. When it splits into two $\P^1$'s, this is the local version of having a one-cycle of the elliptic fiber collapse, hence, this is a remnant of the T-monodromy. We also still see an orientifold monodromy $-\I_2$ that exchanges the two $\P^1$'s. Hence, this component of the fourfold only sees the perturbative subgroup of $SL(2, \Z)$.
This is what makes this limit appropriate for studying F-theory at weak coupling.

\section{A massless ${\rm U}(1)$: brane/image-brane system} \label{sec:brane/imagebrane}
In this section, we will present the simplest example of a ${\rm U}(1)$ gauge symmetry that does not acquire a St\"uckelberg mass.

Let us start from the perturbative picture in IIB on a CY threefold $X_3$, given by a hypersurface of the form:
\begin{equation}
\xi^2 = b_2
\end{equation}
with orientifold involution $\xi \rightarrow -\xi$. The most generic D7-tadpole saturating brane must have the form of a Whitney umbrella \cite{Collinucci:2008pf, Braun:2008ua}:
\begin{equation}
b_4^2-\xi^2\,b_6 = 0\,.
\end{equation}
The easiest way to create a single ${\rm U}(1)$ in IIB is to impose that $b_6$ be a square, i.e. $b_6 = a_3^2$, (equivalently $a_6=0$). This case was studied in \cite{Grimm:2010ez, Berglund:1998va} and dubbed the `${\rm U}(1)$ restriction'. In this case, the total brane splits into a brane and an orientifold image-brane
\begin{equation}
{\rm D7}: (b_4+\xi\,a_3) = 0\,, \quad \,\sigma({\rm D7}):(b_4-\xi\,a_3)=0, 
\end{equation}
such that these lie in the same homology class, i.e. $[D7] = \sigma^*[D7] \in H_4(X_3)$. Here, we expect a ${\rm U}(1)$ gauge group that remains massless at non-zero string coupling $g_s$. 

Let us first analyze the weak coupling limit. From \eqref{WEeq}, we can determine the shape of the `perturbative component' of the central fiber, which we called $W_E$, and write it in the following suggestive form:
\begin{equation}
(y+a_3\,t)\,(y-a_3\,t) = s\,(b_2\,s+2\,b_4\,t)\,.
\end{equation}
This manifold has the classic shape of a conifold, $A\,B = C\,D$. Indeed, there is a curve worth of conifold singularities at the ideal $(y, a_3, s, b_4)$. The curve in question is the ${\rm SU}(2)$-enhanced matter curve where the brane meets its image \emph{outside} the O7-plane, at the ideal $(a_3, b_4)$.

The singularity is easily disposed of via the usual small resolution, by defining two variables $[x_1: x_2]$ of a $\P^1$, and imposing:
\begin{equation}
\left(\begin{array}{cc}y+s & b_4 \\ b_2\,s+2\,b_4\,t & y-a_3\,t \end{array}\right) \left(\begin{array}{c} x_1 \\ x_2 \end{array}\right) = 0\,.
\end{equation}
in the ambient space $Y_5 \times \P^1$ \cite{Braun:2011zm}.

So much for the perturbative part. Now let us add the missing $\lambda s^3$ term to complete the elliptic fiber of the fourfold over then central fiber at $\epsilon = t\,\lambda=0$. We see that the conifold shape persists:
\begin{equation}
(y+a_3\,t)\,(y-a_3\,t) = s\,(\lambda \, s^2+b_2\,s+2\,b_4\,t)\,\quad \cap \quad t\,\lambda = 0\,.
\end{equation}
Now we can easily take this factorizable form outside the central fiber, which means going to strong coupling, and write the fourfold as
\begin{equation}
(y+a_3\,\,z^3\,\epsilon)\,(y-a_3\,z^3\,\epsilon) = s\,(s^2+b_2\,s\,z^2+2\,b_4\,z^4\,\epsilon) \quad \subset X_5=\P_{2,3,1}(\cO_{B_3}\oplus\cO_{B_3} \oplus K )\,.
\end{equation}
Hence, the generic CY fourfold outside Sen's limit still has a codimension three singularity of conifold type that admits a small, K\"ahler resolution. From this we conclude that the ${\rm U}(1)$ does not acquire a St\"uckelberg mass term. 

To construct the corresponding harmonic two-form $\omega_+-\omega_-$ we define the following divisors with ideals:
\begin{equation}
{D_s}_\pm: (y\pm a_3\,z^3\,\epsilon,\quad s)\,, \qquad {\rm and} \quad {D_Q}_\pm: (y\pm a_3\,z^3\,\epsilon,\quad s^2+b_2\,s\,z^2+2\,b_4\,z^4\,\epsilon)\,.
\end{equation}
It can be easily shown that $[{D_s}_+]-[{D_s}_-] = [{D_Q}_-]-[{D_Q}_+]$. Therefore, we can define the ${\rm U}(1)$ via the Poincar\'e dual of $[{D_s}_+]-[{D_s}_-]$

\section{A massive ${\rm U}(1)$}\label{sec:examples2}
We will now present the simplest example of a system that exhibits a ${\rm U}(1)$ gauge symmetry at weak coupling that develops a mass via the St\"uckerberg mechanism, and show that, at strong coupling, the singular geometry only admits a non-K\"ahler resolution.

\subsection{The setup: Two brane/image-brane pairs}

In order to avoid cumbersome D7-tadpole constraints, we will use a simple non-compact model for our geometry: $B_3 \equiv \C^3$. As it turns out, this space will perfectly capture the essential phenomena at play. 

Let $\C^3$ have coordinates $(x_1, x_2, x_3)$. In order to do IIB string theory, we need to define a CY double cover. In this case, we will take 
\begin{equation} \label{nonconifoldthreefold}
X_3: \, \xi^2 = 1+x_1\,x_2\, \subset \, \C^4
\end{equation}
where the ambient $\C^4$ has coordinates $(\xi, x_1, x_2, x_3)$, and $\xi \rightarrow -\xi$ is the orientifold involution. We will put one D7 brane at $x_1=0$, and the other one at $x_2=0$. Each one of these branes actually splits into a brane/image-brane pair with ideals:
\begin{equation}
{\rm D7_i}:\, (x_i,  \xi-1)\,,\qquad {\rm D7_i'}:\, (x_i, \xi+1)\,,\quad {\rm for }\quad i=1,2\,.
\end{equation}
Since $X_3$ is smooth, these divisors are Cartier, and hence each can \emph{locally} be defined by a single equation. For instance, if we take the patch $U$ where $\xi+1 \neq 0$, then we can rewrite the equation for the threefold as:
\begin{equation}
X_3|_U: \quad \xi-1 = \frac{x_1\,x_2}{\xi+1}
\end{equation}
and define the $D7_1$ by the single equation $x_1=0$.
However, \emph{globally}, we cannot single out the $D7_1$ from its image by a single equation. We always need two equations: $(\xi-1,x_1)$. This simple fact is a local remnant of the fact that, in a compact setting, the $D7_1$ and its image lie in different homology classes. To be clear, our claim is the following: Because the $D7_1$ and the $D7_1'$ cannot separately be defined globally by one equation intersected with \eqref{nonconifoldthreefold}, then, in a compactification of this model, $[D7] \neq \sigma^*[D7] \in H_4(X_3)$. See appendix \ref{sec:appendix} for proof of this claim.
Since the same applies to the $D7_2$, we expect a ${\rm U}(1) \times {\rm U}(1)$ gauge group that will become massive at non-zero $g_s$.

Let us start at weak coupling by looking at $W_E$. In this case, we have 
\begin{equation}
b_2 = 1+x_1\,x_2\,, \quad b_4 = x_1\,x_2\,, \quad {\rm and} \quad b_6 = x_1\,x_2\,,
\end{equation}
such that $\Delta_E \equiv b_4^2-b_2\,b_6 = -x_1\,x_2$. Then, $W_E$ takes the simple form:
\begin{equation}
(y+s)\,(y-s) = x_1\,x_2\,(s+t)^2\,.
\end{equation}
The singular locus is at the ideal $(y, s, x_1, x_2)$. Indeed, since each brane in this model intersects its image only at the O7-plane, there is no charged matter, as there is no antisymmetric ${\rm U}(1)$ matter. Hence, we can only detect the relative linear combination ${\rm U}(1)_{2-1}$ at the ${\rm SU}(2)$-enhanced matter curve given by $(x_1, x_2)$. This curve carries the bifundamental matter charged under this relative group. 
In this weak coupling limit, we see that this curve worth of conifold singularities admits a small resolution, as it has the form $AB = CD$.
This is consistent with the fact that, in the limit $g_s\rightarrow 0$, the expected mass goes to zero: at $\epsilon=0$, there are two independent six-cycles in $W_E$, $(y\pm s,x_1)$ and $(y\pm s,x_2)$, that generate two massless ${\rm U}(1)$'s by expanding $C_3$.

Now, to move away from the weak coupling limit, we simply add the missing $\lambda\,s^3$ term, and then export this out of the central fiber. This amounts to setting the hypersurface equation to:
\begin{equation} \label{nonkaehler4fold}
(y+s\,z)\,(y-s\,z) = x_1\,x_2\,z^2\,(s+z\,^2\epsilon)^2+s^3\,.
\end{equation}
The fourfold is still singular at $(y, s, x_1, x_2)$. So, in this sense, we still detect the ${\rm U}(1)_{2-1}$. However, we have lost our nice $AB=CD$ form. In fact, we will now prove that this curve worth of conifold singularities only admits a \emph{non-K\"ahler} small resolution\footnote{It also admits a large K\"ahler resolution, which necessarily breaks the CY condition.}.
Moreover, due to the lost of the factorization $AB=CD$ at finite $\epsilon$, the six-cycles $(y\pm s,x_1)$ and $(y\pm s,x_2)$ disappear from $H_4(X_4,\mathbb{Z})$ and the corresponding ${\rm U}(1)$'s are no more massless.

\subsection{Non-K\"ahler resolution}
In this subsection, we will show that the relative ${\rm U}(1)_{2-1}$, which is expected to acquire a St\"uckelberg mass, gives rise to a fibration of conifolds over the matter curve $(x_1, x_2)$, where the two different branes species meet, such that the singularity only admits a \emph{non-K\"ahler} small resolution.

Let us first simplify the form of our fourfold \eqref{nonkaehler4fold}. The only singularity is at $(y, s, x_1, x_2)$. We can therefore restrict to the patch where $s+\epsilon\,z^2 \neq 0$, and set that factor to one, since $(y, s, z)$ form a $\P^2_{3,2,1}$. Now our singularity is simply:
\begin{equation} \label{conifoldpluscube}
y^2-s^2 = x_1\,x_2+s^3\,,
\end{equation}
whereby the cubic term spoils the typical resolvable conifold shape, but does not actually affect the singularity, since it is of higher order.

In order to see how the non-K\"ahler resolution comes about, let us first look at the standard conifold from a slightly different perspective. Consider the same hypersurface as before, but dropping the cubic term, and suppressing the coordinate $x_3$:
\begin{equation} \label{standardconifold}
y^2-x_1\,x_2 = s^2 \quad \subset \quad \C^4\,, \quad \text{with coordinates} \quad (y,s,x_1,x_2)
\end{equation}
We can view this threefold as a fibration of deformed $A_1$-singularities over the $s$-plane, whereby only the central fiber at $s=0$ is singular. The two inequivalent small resolutions consist in blowing up the threefold at the ideals $(y\pm s, x_1)$. After doing this, we still have a fibration of deformed $A_1$-surfaces over the $s$-plane, but now the central fiber has a resolved $A_1$-surface. As is well-known, and can be deduced from the hyper-K\"ahler structure of K3 surfaces, resolving is equivalent to deforming for ADE surface singularities.

At any rate, this new threefold has exactly one compact two-cycle: In the central fiber, it is given by the exceptional $\P^1$. Outside the central fiber, it is described by the non-holomorphic sphere given by the real slice of \eqref{standardconifold}:
\begin{equation}
\Re(y)^2+\Re(x_1-x_2)^2+\Re(x_1+x_2)^2 = \Re(s)^2 \quad \subset \quad \R^3\,,
\end{equation}
where the $\R^3$ has coordinates $(\Re(y), \Re(x_1-x_2), \Re(x_1+x_2))$ and $\Re(s)$ is the radius of the sphere.
Since the resolved conifold is K\"ahler, we can compute the volume of the exceptional $\P^1$ by integrating the K\"ahler form $J$ on it. If this $\P^1$ \emph{were} the boundary of a three-chain $\Sigma_3$, then Stokes theorem would imply that
\begin{equation}
{\rm Vol}(\P^1) = \int_{\P^1} J = \int_{\Sigma_3} dJ \neq 0\,,
\end{equation}
which would imply that the K\"ahler form is not closed, contradicting the assumption that the resolution is K\"ahler. 

How do we prove that $\P^1 \neq \partial \Sigma_3$ for some $\Sigma_3$? In this situation, any such three-chain would be shaped like a family of spheres, ending on the $\P^1$ at one extreme, and pinching off somewhere, like the tip of a cigar. However, in this geometry, the family of spheres extends over the whole $s$-plane from the origin all the way to infinity, and never pinches off. Hence, any three-chain would always have at least two boundaries: one at the origin, and one at infinity. This invalidates the use of Stokes theorem as above.

Now we are ready to face our singularity from \eqref{conifoldpluscube}. First, we take a small neighborhood of the singularity, such that we can neglect the cubic term. There, we recover the standard form of the conifold, which we can resolve. Since birational tranformations are local operations, we can always resolve in one neighborhood, and then patch the geometry together.
So, locally, it looks like we can again claim that there are no three-chains bounding the exceptional $\P^1$. However, if we again look at our geometry as a fibration of deformed $A_1$-singularities over the $s$-plane:
\begin{equation}
s^2\,(s+1) = y^2-x_1\,x_2\,,
\end{equation}
we see that, at the locus $s=-1$, the $A_1$-fiber is singular again. This means that the non-holomorphic two-sphere collapses over this point, even though we are not at a singular point of the threefold. Therefore, one can define a three-chain $\Sigma_3$ as the family of spheres over the interval $s \in [1,0]$, such that $\partial \Sigma_3 = \P^1$ at $s=0$. Therefore, by Stokes' theorem, we have shown that $dJ \neq 0$.

To apply this to our fourfold, all we do is fiber everything over the curve parametrized by $x_3$. Over each point of this curve lies a $\P^1$ that is a boundary. Hence, the fourfold has been resolved to a non-K\"ahler manifold.

\bigskip

We are now ready to make our general claim, which was anticipated in the introduction:

${\rm U}(1)$ gauge symmetries manifest themselves as fibrations of conifold singularities over matter curves in IIB, provided these curves host matter charged under the ${\rm U}(1)$. 
In the weak coupling limit of Clingher, Donagi and Wijnholt, these singularities will always admit a small, K\"ahler resolution.
If the ${\rm U}(1)$ remains massless at strong coupling, then the full-fledged CY fourfold admits a small, K\"ahler resolution. If it develops a St\"uckelberg mass term, then it will only admit a small, but non-K\"ahler resolution (as conjectured in \cite{Grimm:2011tb}).

We will see by way of examples that this structure prevails in various brane setups. 

\section{Globally defined examples}\label{sec:globalexamples}
In this section, we move on to more intricate global models: One with a massive ${\rm U}(1)$, a related model with a massless ${\rm U}(1)$, and finally a model with one massive and one massless ${\rm U}(1)$.

\subsection{A massive ${\rm U}(1)$} 
We can easily generalize our previous example to a setup with one brane/image-brane pair carrying a ${\rm U}(1)$ that acquires mass, and a Whitney brane that has no gauge group, but simply saturates the D7-tadpole. We can now work in generality over compact manifolds.

Let $X_3$ be a CY threefold given by a hypersurface of the form:
\begin{equation}
\xi^2 = b_2 \equiv a_1^2+\sigma\,\tilde a_2\,,
\end{equation}
where $a_2 = \sigma \, \tilde a_2$, and $\sigma=0$ is the locus of the brane/image-brane pair. Note that $X_3$ will have conifold singularities at the locus given by the ideal $(\xi, a_1, \sigma, \tilde a_2)$ that cannot be resolved crepantly without breaking the orientifold involution \cite{Donagi:2009ra}. It is also possible to have models where these points are excised from the geometry \cite{Collinucci:2009uh, Krause:2012yh}.

We place a split $I_1$ brane at $\sigma=0$, and an orientifold invariant brane at $\sigma\,\eta^2-b_2\,\chi$, for some $\sigma$ and $\chi$ of appriopriate degree, such that
\begin{equation}
\Delta_E = b_4^2 - b_2\, b_6 = \sigma \,(\sigma\,\eta^2-b_2\,\chi)\,.
\end{equation}
This means we are setting $b_4 = \sigma\,\eta$ and $b_6 = \sigma\, \chi$. Putting this into the equation for the perturbative part of the CY fourfold $W_E$ \eqref{WEeq}, we get the following form:
\begin{equation}
(y+a_1\,s)\, (y-a_1\,s) = \sigma\,(\tilde a_2\,s^2+2\,\eta\,s\,t+\chi\,t^2)\,.
\end{equation}
This fourfold has a curve worth of conifold singularities at the ideal $(y, s, \sigma, \chi)$. The curve in question is the matter curve $(\sigma, \chi)$, where the two brane systems intersect. 

The two branes also intersect in a non-generic way over a curve in the O7-plane at $(\sigma, b_2)$. Naively, one would expect an $SO(4)$ enhancement from the IIB perspective. However, such a group is missing from the Tate classification of singularities. We expect that the non-perturbative splitting of the O7-plane will prevent such an enhancement. Indeed, the corresponding singularities at $(y, a_1, \sigma, \tilde a_2\,s^2+2\,\eta\,t\,s+\chi\,t^2)$ are washed away at strong coupling. This form of quantum splitting has been studied in several cases in \cite{Cvetic:2010rq}. 

The fourfold has the classic $AB = CD$ shape that admits a small, K\"ahler resolution, so we conclude that there is a ${\rm U}(1)$ gauge group. Indeed, the brane at $\sigma$ splits into a brane/image-brane  $D7_\sigma/D7_\sigma'$ pair at $(\sigma, \xi\pm a1)$.

These two branes do not lie in the same class in $H_4(X_3)$, see appendix \ref{sec:appendix}. Hence, we expect the ${\rm U}(1)$ to develop a St\"uckelberg mass. By going to the strong coupling regime, the fourfold now takes the form\footnote{To avoid cluttering, we suppress factors of $z$, since all singularities are located away from $z=0$.}:
\begin{equation}
(y+a_1\,s)\, (y-a_1\,s) = s^3+\sigma\,(\tilde a_2\,s^2+2\,\eta\,s\,\epsilon+\chi\,\epsilon^2)\,,
\end{equation}
which is still singular at $(y, s, \sigma, \chi)$. Note, however, that we have lost the ability to perform a small, K\"ahler resolution. However, a small, non-K\"ahler resolution is possible, just as it the previous example. This is the F-theoretic manifestation of the St\"uckelberg mass.

\subsection{A massless ${\rm U}(1)$}
Let us now move away from the split $I_1$ Ansatz by allowing the threefold to have a generic form 
\begin{equation}
\xi^2 = b_2\,, \quad \text{for generic} \quad b_2\,,
\end{equation} 
and take a system with a brane/image-brane pair, where both branes are homologous, plus a so-called Whitney brane. For simplicity, let us take the brane/image-brane pair such that it doesn't self-intersect away from the O7-plane:
\begin{equation}
b_4^2-b_2\,b_6 = (\eta_1^2-\xi^2) \, (\eta_2^2-\xi^2\,\chi_2)\,.
\end{equation}
We see that the first factor splits into $(\eta_1\pm \xi)$, which gives rise to a ${\rm U}(1)$. In fact, this case is akin to the so-called ${\rm U}(1)$-restriction of \cite{Grimm:2010ez}, except that the split brane is not alone. The second factor is orientifold invariant, and has a $\Z_2$ gauge group \cite{Collinucci:2008pf}. 
This system is achieved by choosing:
\begin{equation}
b_4 = \eta_1\,\eta_2\,, \quad b_6 = \eta_1^2\,\chi_2+\eta_2^2-b_2\,\chi_2\,.
\end{equation}
Plugging this Ansatz into $W_E$ \eqref{WEeq}, we get
\begin{equation}
y^2 = b_2\,s^2+2\,\eta_1\,\eta_2\,s\,t+(\eta_1^2\,\chi_2+\eta_2^2-b_2\,\chi_2)\,t^2\,,
\end{equation}
which can be re-written in the following suggestive form:
\begin{equation}
(y+\eta_2\,t+\eta_1\,s)\,(y-\eta_2\,t-\eta_1\,s) = (s^2-\chi_2\,t^2)\,(b_2\,-\eta_1^2)\,.
\end{equation}
Here, we clearly recognize a family of conifold singularities. Note, however, that the base of this family isn't the matter curve in $B_3$ given by the ideal $(\eta_1^2-b_2, \eta_2^2-b_2\,\chi_2)$, but instead over a $(\eta_1^2-b_2, s^2-\chi_2\,t^2)$. The first equation clearly describes the brane/image-brane pair. The second one seems to describe a blown-up version of the Whitney brane.

Since the singularity has the standard factorizable form, we can perform a small, K\"ahler resolution on it, thereby detecting the ${\rm U}(1)$ gauge group. Based on the analysis of 
\cite{Grimm:2011tb}, we expect this ${\rm U}(1)$ to remain massless at strong coupling. However, if we add the $s^3$ term to this model and write the non-perturbative F-theory fourfold
\begin{equation}
(y+\eta_2\,\epsilon+\eta_1\,s)\,(y-\eta_2\,\epsilon-\eta_1\,s) = (s^2-\chi_2\,\epsilon^2)\,(b_2\,-\eta_1^2)+s^3\,,
\end{equation}
we see that we have spoiled the usual form that admits a small, K\"ahler resolution.\footnote{Moreover the manifold does not develop a conifold singularity along a curve, but just a point singularity. This deformation is not preserving the matter content.} It would thus seem that we have run into a contradiction. The way out of this paradox is to realize that there is more than one way of exiting the central fiber in this family of CY fourfolds. Alternatively, one might say that there are different ways of reaching a weak coupling limit, as was explored for instance in \cite{Aluffi:2009tm, Esole:2012tf}. In this particular case, instead of simply adding $s^3$, we should add $s\,(s^2-\chi_2\,\epsilon^2)$, yielding
\begin{equation}
(y+\eta_2\,\epsilon+\eta_1\,s)\,(y-\eta_2\,\epsilon-\eta_1\,s) = (s^2-\chi_2\,\epsilon^2)\,(s+b_2\,-\eta_1^2)\,.
\end{equation}
Now our non-perturbative, full-fledged F-theory fourfold still has the family of conifold singularities, but admits a small, K\"ahler resolution.

\subsubsection*{Variation}

For a slight variation of this model, let us see what happens if we choose $\chi_2 \equiv \psi_2^2$. Then this model will exhibit three small-resolvable conifold curves, only two of which are independent:
\begin{equation}
(y+\eta_2\,\epsilon+\eta_1\,s)\,(y-\eta_2\,\epsilon-\eta_1\,s) = (s+b_2\,-\eta_1^2)\,(s+\psi_2\,\epsilon)\,(s-\psi_2\,\epsilon)\,.
\end{equation}
This correlates perfectly with our expectations from IIB theory: 

\begin{itemize}
\item We have two brane/image-brane pairs. The first one meets its own orientifold image only on the O7-plane. Hence, there is no matter charged under its diagonal ${\rm U}(1)_1$. 

\item The second one does have an ${\rm SU}(2)$-enhancement when it intersects its own image at the ideal $(\eta_2, \psi_2)$, which lies outside the O7-plane. From this we expect matter charged under the symmetric representation of ${\rm U}(1)_2$. Since in this case, $[D7_2] = \sigma^*[D7_2]$, we expect this group to remain massless.
In the fourfold, we see the associated two-form for this ${\rm U}(1)_2$ as the Poincar\'e dual of $D_+-D_-$ for
\begin{equation}
D_\pm: \quad (y\pm(\eta_2\,\epsilon+\eta_1\,s)\,, \, s+\psi_2\,\epsilon)\,.
\end{equation}

\item Finally, the matter curve where the two brane systems intersect, contains matter charged under the bifundamental group ${\rm U}(1)_{2-1}$, which we see via the divisors:
\begin{equation}
D_\pm: \quad (y\pm(\eta_2\,\epsilon+\eta_1\,s)\,, \, s+b_2\,-\eta_1^2)\,.
\end{equation}

\end{itemize}

\subsection{${\rm U}(1)_{\rm massless} \times {\rm U}(1)_{\rm massive}$}
Here we will see the most intricate example so far: Two brane/image-brane systems such that $[D7_i] \neq \sigma^*[D7_i]$ for $i=1,2$. However, the $D7_1$ meets its image only on the O7-plane, whereas the $D7_2$ meets its image along an ${\rm SU}(2)$-enhancements.
We expect the corresponding (diagonal) ${\rm U}(1)$'s to get non-zero masses through the geometric St\"uckelberg mechanism. 

However, by inspecting the corresponding axionic coupling in the type IIB effective action one obtains that there is one massless combination of these two ${\rm U}(1)$'s. This massless ${\rm U}(1)$ should be visible also away from the weak coupling limit. 

Let us choose the $b_i$ such that we have two branes plus their images in the weak coupling type IIB setup:
\begin{equation}
 b_2 = a_1^2+ \tilde{a}_2\sigma\,, \qquad b_4 = a_1a_3 + \tilde{a}_{4}\sigma\,, \qquad b_6=a_3^2 \:.
\end{equation}
Correspondingly $\Delta_E= \sigma\,( 2 a_1a_3\tilde{a}_{4} - \tilde{a}_{2}a_3^2 +\sigma\,\tilde{a}_{4}^2 )$.
The Calabi-Yau threefold is of the type necessary for having split ${\rm U}(1)$'s:
\begin{equation}
X_3\,:\,\, \xi^2 = a_1^2+\sigma\,\tilde{a}_{2} 
\end{equation}
The D7-locus $D7_\sigma$ at $\sigma=0$ splits into the two branes at $\xi-a_1=0$ and $\xi+a_1=0$. The locus $D7_D$: $2 a_1a_3\tilde{a}_{4} - \tilde{a}_{2}a_3^2 +\sigma\,\tilde{a}_{4}^2=0$ also splits into a brane and its image when it is intersected with the $X_3$ equation, even though it is not manifest \cite{Krause:2012yh}. In order to see this, we work over the function field of $X_3$, and simply divide the equation by $\sigma$, yielding:
\begin{equation}
\tfrac{1}{\sigma}\,(2 a_1a_3\tilde{a}_{4} - \tilde{a}_{2}a_3^2 +\sigma\,\tilde{a}_{4}^2) = (a_3+\tilde{a}_4\,\sigma+\xi\,a_3)\,(a_3+\tilde{a}_4\,\sigma-\xi\,a_3)\,,
\end{equation}
which is clearly reducible.

The fourfold $W_E$ can be written as:
\begin{equation}
W_E \,: \qquad 
(y-s\,a_1-t\,a_3)(y+s\,a_1+t\,a_3) = \,\sigma\,s\,(\tilde{a}_{2}s+2\tilde{a}_{4}t) 
\end{equation}
The manifold has the form $AB=CDE$ and correspondingly has a number of conifold singularities. In particular they are at the ideals $(y,a_3,s,\tilde{a}_4)$, $(y,a_3,s,\sigma)$ and $(y,a_1s+a_3t,\tilde{a}_2s+2\tilde{a}_4t,\sigma)$. From a  perturbative type IIB analysis one can compute what is the number of massless ${\rm U}(1)$'s in this configuration. Both loci $D7_\sigma$ and $D7_D$ are made up of a brane and its image in different homology classes in $X_3$. Hence, as explained at the beggining, after a suitable normalization, we expect the linear combination ${\rm U}(1)_\sigma+{\rm U}(1)_D$ to remain massless at strong coupling.

In F-theory, the massless ${\rm U}(1)$'s are related to the new non-Cartier divisors that arise when we have the factorized conifold singularity. In  this example, the weak coupling fourfold $W_E$ has the particular factorized form $AB=CDE$. This implies two independent such cycles (compatible with 4d Poincar\'e invariance), that we can choose to be:
\begin{eqnarray}
\omega^\sigma_+ - \omega^\sigma_-   &\mbox{ with }& \omega^\sigma_\pm\,:\,\,\,  y\pm (a_1s+a_3t)=\sigma=0 \,, \\  
\omega^D_+ - \omega^D_-   &\mbox{ with }&  \omega^D_\pm\,:\,\,\,y\pm (a_1s+a_3t)=\tilde{a}_{2}s+2\tilde{a}_{4}t=0  \:.
\end{eqnarray}
These are the six-cycles related to the two ${\rm U}(1)$'s that are indeed massless at $g_s\rightarrow 0$.
We expect that a combination of the six-cycles survive the deformation away from weak coupling. 
It is not difficult to understand what is happening, by looking at the equation of the fourfold away from weak coupling. It can be descibed as
\begin{equation}
(y-s\,a_1\,z-\epsilon\,a_3\,z^3)(y+s\,a_1\,z+\epsilon\,a_3\,z^3) = s \left( s^2 + \sigma\,(\tilde{a}_{2}s\,z^2+2\tilde{a}_{4}\,\epsilon\,z^4) \right)\:.
\end{equation}
In this case, we see only one independent  non-Cartier divisor, that we would like to associate to the massless ${\rm U}(1)$ combination:
\begin{eqnarray}
 \omega_+ - \omega_-   &\mbox{ with }&  \omega_\pm \,:\,\,\, y\pm (a_1s\,z + a_3\,\epsilon\,z^3)= s^2 + \sigma\,(\tilde{a}_{2}s\,z^2+2\tilde{a}_{4}\,\epsilon\,z^4)  =0 \:.
\end{eqnarray}
This interpretation is supported by the fact that in the weak coupling limit, the six-cycles $\omega_\pm$ split into $\omega^\sigma_\pm+\omega^D_\pm$. Away from weak coupling, the fourfold has still three curves of conifold singularities, but only two of them admit a small resolution. The third one, related to the massive combination, should be resolved by a non-K\"ahler resolution.

\section{Discussion}
In this paper, we have provided a geometric picture for understanding Abelian gauge symmetries, and discriminating massive from massless ones. We find that all ${\rm U}(1)$'s that do not decouple, i.e. such that there is matter charged under them, manifest themselves as one-parameter families of conifold singularities of the perturbative fourfold $W_E$, admitting small, K\"ahler resolutions. Codimension three singularities admit such resolutions provided there is a non-Cartier divisor.

The simplest example is when $W_E$ is a hypersurface of the form $A\,B = C\,D$. Here, a non-Cartier divisor would be the ideal $(A, C)$. However, it may happen that this form is not manifest. For instance, one might discover that, although $W_E$ does not have this form, it admits non-Cartier divisor that is simply not a complete intersection. However, by working over the function field of $B_3$, one should still be able to bring the fourfold to the $A\,B = C\,D$ form.

At strong coupling, the ${\rm U}(1)$ will remain massless if and only if the full-fledged fourfold retains the form $A\,B = C\,D$. This means  concretely, that one should be able to add a polynomial of leading term $s^3$ without spoiling the form. Therefore, if $W_E:  A\,B = C\,D$ with $A$ a monic polynomial in $s$ of degree $d \leq 2$, then we can always add $s^{3-d}\,A$, such that the Weierstrass equation will be
\begin{equation}
W: \quad A\,(B+s^{3-d}) = C\,D\,,
\end{equation}
which still admits a small, K\"ahler resolution.

In this paper, we have constructed setups with purely Abelian gauge groups, in order to convey our message most efficiently. When dealing with non-Abelian singularities, one must beware of the following subtlety: Take a stack $D7_G$, with an ${\rm U}(N)$ gauge group, such that the diagonal ${\rm U}(1)$ is expected to be massive, and a flavor stack $D7_f$. By resolving the codimension \emph{two} singularity over the 7-brane, one notices that the codimension \emph{three} singularity over the enhanced matter curves $(D7_G, D7_f)$ is automatically resolved. This would appear to contradict our claim that such singularities do not admit crepant, K\"ahler resolutions. 
From the IIA perspective, such a Cartan resolution corresponds to separating two of the D6-branes, $D6_1$ and $D6_2$ of the stack $D6_G$ along the T-duality circle. The enhancement we see represents the fact that, while these two branes are also being separated about the origin of the coordinate system, the `flavor' brane $D6_f$ is staying put at the origin. By fibering the M-theory circle over the intervals $[D6_1, D6_f]$ and $[D6_f, D6_2]$, we see two $\P^1$'s.
From the perspective of IIB on $\R^3 \times S^1$, we are switching on a Wilson line $A_2-A_1$, along the Cartan of the gauge group. This necessarily breaks the bifundamental ${\rm U}(1)$ between the gauge and the `flavor' stack, even though we are forced to keep $A_f=0$ due to its mass. 
Resolving the codimension \emph{three} singularity, on the other hand, corresponds to displacing the $D6_f$ relative to the $D6_G$ stack, or, alternatively, to switching on a Wilson line along $A_f$. Hence, our claim is unaffected by this subtlety. 

Finally, our results provide a starting point to describe massive ${\rm U}(1)$'s explicitly in F-theory. An interesting direction to explore would be the construction of fluxes along such massive gauge groups, by inspecting how the harmonic two-form that detects the massive ${\rm U}(1)$ at $\epsilon=0$ can be extended to a non-harmonic two-form in the fourfold at strong coupling.

\section*{Acknowledgements}

We have benefitted from discussions with Robert Richter, Raffaele Savelli and Taizan Watari.
A. C. is a Research Associate of the Fonds de la Recherche Scientifique F.N.R.S. (Belgium).
The work of A.P.B is supported by the STFC under grant ST/J002798/1.

\appendix

\section{Appendix: Rigidity implies $[D7] \neq \sigma^*[D7]$} \label{sec:appendix}
The most easily conceived orientifold involutions on a CY threefold $X_3$ typically act trivially on the even homologies $H_{2\,*}(X_3)$, so that $h^{1,1}_-=0$. However, many interesting cases  with $h^{1,1}_- \neq 0$ have been explored, for instance in \cite{Blumenhagen:2008zz, Collinucci:2009uh, Blumenhagen:2009up}. In these models, it is possible for a non-invariant D7-brane to have an orientifold image which is in a different homology class. In such cases, the associated ${\rm U}(1)$ gauge groups acquire St\"uckelberg mass terms, as explained in \cite{Grimm:2011tb}.

Let $\sigma$ be a holomorphic involution giving rise to O7/O3-planes. There will necessarily be a single coordinate or polynomial, call it $\xi$, such that $\sigma: \xi \mapsto -\xi$. Then, a typical non-invariant divisor will have the form $\eta+\xi\,\psi=0$, such that its image lies at $\eta-\xi\,\psi=0$. However, since both equations can be seen as deformations of each other, the respective divisors will lie in the same homology class.
The only way to achieve a divisor/image-divisor pair $D/ \sigma(D)$, such that $[D] \neq \sigma^*[D] \in H_2(X_3)$, is by considering non-complete intersections with $X_3$. The easiest conceivable model is constructed by imposing that $X_3$ be given by a hypersurface of the form:
\begin{equation} \label{cy3twosigmas}
\xi^2 = a_1^2+\sigma_1\, \sigma_2
\end{equation}
where $(a_1, \sigma_1, \sigma_2)$ are polynomials. In this case, the divisors $\sigma_i = 0$ for $i=1,2$ are each reducible, with their two components at 
\begin{equation}
D_i:\quad (\sigma_i, \xi+a_1)\,, \qquad \sigma(D_i):\quad (\sigma_i, \xi-a_1)\,.
\end{equation}

Let us assume that the locus $(\xi, a_1, \sigma_1, \sigma_2)$ is empty so that $X_3$ is smooth. In the language of \cite{Krause:2012yh}, that corresponds to excising $E_6$-points in ${\rm SU}(5)$-models. Now both divisors are Cartier, so they are locally cut out by a single equation inside $X_3$, but not globally. In what follows, we will now prove that $[D_i] \neq \sigma^*[D_i]$ in three steps.

\subsubsection*{1) If $[D_i]$ is rigid $\Rightarrow [D_i] \neq  \sigma^*[D_i]$}

The first we prove that if least one of the $D_i$ is rigid, then $[D_i] \neq \sigma^*[D_i]$ for both $i=1, 2$. To each divisor class $[D_i]$ is associate a line bundle $\cL_i$, and each representative is given by the zero locus of a section $s_i$. Every line bundle admits a `zero section' $s \equiv 0$, but it may or may not admit non-trivial holomorphic sections. Suppose that $D_i$ and $\sigma(D_i)$ are in the same divisor class. Then they are associated to two different sections $s_a, s_b$ of the same line bundle $\cL_i$. But then, one can construct a one-parameter family of section and hence divisors by considering linear combinations $a\,s_a+b\,s_b=0$, where $(a,b) \sim (\lambda\,a,\lambda\,b)$ define a $\P^1$. Therefore, $D_i$ cannot be rigid.

\subsubsection*{2) If, say $[D_1] \neq \sigma^*[D_1]$ is rigid, then $[D_2] \neq \sigma^*[D_2]$}

From the defining equation of $X_3$, $\xi^2 = a_1^2+\sigma_1\,\sigma_2$, we can easily deduce some equalities. Since
\begin{equation}
\xi+a_1 = 0 \quad \cap \quad X_3 \qquad \cong \qquad \xi+a_1 = 0 \quad \cap \quad \sigma_1\,\sigma_2=0\,,
\end{equation}
the class $[\xi+a_1] = [D_1]+[D_2] \subset H_4(X_3)$. On the other hand, the two polynomials $\xi\pm a_1$ are sections of the same line bundle $\bar K$. Therefore, we have
\begin{equation}
[\xi+a_1] = [\xi-a_1] \quad \Rightarrow \quad [D_1]+[D_2]  = \sigma^*[D_1]+\sigma^*[D_2]\,.
\end{equation}
Therefore, $[D_1] \neq \sigma^*[D_1] \Rightarrow [D_2] \neq \sigma^*[D_2]$, as claimed.
Now, all we need to prove is that at least one of the $D_i$ is rigid.

\subsubsection*{3) At least one of the $D_i$ is rigid}

Suppose for simplicity\footnote{The case $[D_1] = [D_2]$ is more subtle, but the answer will be the same. For simplicity, we will focus on the more generic case.} that $[D_1] \neq [D_2]$. Define $\cL_{\sigma_i}$ as the line bundle corresponding to $\sigma_i$. Note that $\cL_{\sigma_i} = \cL_i \otimes \sigma^*(\cL_i)$. Let $\xi_\pm = \xi\pm a_1$. Then, the respective equations for the divisors can be written in matrix form:
\begin{equation}
D_1:\quad \begin{pmatrix} \xi_+& \sigma_2 \\ \sigma_1 & \xi_- \end{pmatrix} \begin{pmatrix} 1 \\ \delta_1 \end{pmatrix} = 0\,, \qquad D_2:\quad \begin{pmatrix} \xi_+& \sigma_1 \\ \sigma_2 & \xi_- \end{pmatrix} \begin{pmatrix} 1 \\ \delta_2 \end{pmatrix} = 0\,,
\end{equation}
where $\delta_i \in \Gamma(K \otimes \cL_{\sigma_i})$. These $\delta_i$ are deformation parameters that allow the divisors $D_i$ so move in families. The particular case we started with had $\delta_i \equiv 0$, but if $K \otimes \cL_{\sigma_i}$ admits a non-zero section, then $D_i$ will not be rigid. However, by equation \eqref{cy3twosigmas}, we have the identity $\cL_{\sigma_1} \otimes \cL_{\sigma_2} = \bar K^{2}$, which implies $K \otimes \cL_{\sigma_1} = (K \otimes \cL_{\sigma_2})^{-1}$. This means that, at most, only one of the two $\delta_i$ can be non-zero. For instance, if $K \otimes \cL_{\sigma_1}$ has a non-zero section, then its dual line bundle cannot have one. Therefore, at least one of the two $D_i$ is rigid.

\bibliography{u1.bib}
\bibliographystyle{utphys}
\end{document}